\documentclass[fleqn,twoside]{article}

\usepackage{espcrc2}
\usepackage{graphicx}
\usepackage[figuresright]{rotating}
\newcommand{\beq}{\begin{equation}} 
\newcommand{\eeq}{\end{equation}\noindent}
\title{Neutrino-nucleus interactions: open questions and future projects}

\author{Cristina Volpe\address[IPNO]{Institut de Physique Nucl\'eaire,\\ 
        F-91406 Orsay cedex, France}}
       
\begin{document}

\begin{abstract}
We discuss various issues concerning 
the interactions of nuclei with neutrinos of low impinging energies
(i.e. having several tens of MeV to a few hundred MeV) 
of interest for particle physics, nuclear physics and astrophysics.
We focus, in particular, on open questions as well as possible strategies
to obtain more experimental information. The option of a low-energy beta-beam 
facility is extensively discussed. We also mention its potential
concerning the neutrino magnetic moment.
\vspace{1pc}
\end{abstract}

\maketitle

\section{Introduction}
Neutrino-nucleus interactions are a topic of current great interest.
The motivations for studying such reactions come from 
the necessity of knowing precisely neutrino detector
response since nuclei are often used as neutrino detectors  
e.g. in solar experiments, for supernova observatories or in oscillation
measurements.  Another important motivation
comes from astrophysics and, in particular, from 
understanding the nucleosynthesis of heavy elements during the r-process
\cite{gail,qian,goriely,bahanunucl,bahaastro,terasawa}
or from neutrino-nucleosynthesis
\cite{haxton,woosley,haxtonrec,pinedo}. 

The present knowledge of $\nu$-nucleus interactions exploits the
knowledge of the weak interaction on one hand and the most developed
techniques for describing the nucleus on the other hand.
The models employed take advantage of a wealth of indirect 
as well as direct experimental information. The latter represent, 
however, a limited ensemble of data still. Useful indirect information is provided by related
processes like beta-decay, muon capture, electron scattering or charge-exchange reactions. Some model-independent sum-rules also help in constraining 
the calculations. The direct measurements include  
one measurement on deuteron \cite{deut} and iron \cite{iron} 
and a series a measurements on carbon \cite{expc12} with neutrinos produced by
the decay-at-rest of muons or by the decay-in-flight of pions.

By performing systematic neutrino-nucleus interactions studies
one could address the numerous open questions in this field
\cite{bahanunucl,bahaastro,revuekubo,jpg}.
First, a very precise knowledge of the interactions on some
nuclei, which are currently used as neutrino detectors, is needed.
Here we mention a few.
The best known case is the neutrino-deuteron reaction, relevant for the
SNO experiment \cite{sno}, 
where theoretical calculations reach the few percent precision 
\cite{kubodera}. However, there is still an important unknown quantity $L_{1,A}$
whose determination would improve our knowledge of the {\it pp} reaction in the Sun.
The most studied case, namely neutrino interactions on carbon, still suffer
from a discrepancy between experiment and theory, in particular for the
neutrinos produced from the decay-in-flight of pions, in spite of the efforts
done during several years \cite{c12}. A precise measurement of the reactions
on oxygen is of great interest also in view of the next-generation experiments
involving Megaton detectors like Hyper-K or UNO \cite{uno}. In particular,
their use with the aim of studying CP violation in the lepton sector would
need a very precise determination of the reaction cross sections 
in the range of several hundred MeV. 
Lead represent an interesting
nucleus as well, e.g. for supernova observatories 
\cite{fuller,elliott,volpelead,nuPb,nuFe}.
In this context, the precise measurement of the differential cross sections of
electrons emitted in charge-current events would be very useful, since their measurement
would give us precious information on the neutrino temperatures at emission, if
a core-collapse supernova explodes \cite{nuPb}. 

One of the richnesses of this field is that for increasing  neutrino energies, 
neutrinos probe the nuclear to the nucleon degrees of freedom. 
While the description of the low-energy regime exploits approaches like
the Elementary Particle Theory, Effective Field Theories, or microscopic
models,
such as the shell model or the Random-Phase-Approximation, the Fermi Gas 
is the basis for the theoretical description in the high energy regime. 
In particular, neutrinos can be used to perform nuclear structure studies
since  neutrinos probe nuclear excited states for which little or no 
experimental information is available  (Figure 1) \cite{volpelead}. 
The role of some of these states
in astrophysical contexts is outlined in several papers
\cite{gail,fuller,volpelead,kolbe,jon}.
A larger ensemble of experimental data would put the interpolation, of the 
neutrino-nucleus 
interaction modeling between these two regimes, 
on even more solid grounds. This would also help the extrapolation to the case 
of neutron-rich
nuclei of astrophysical interest that are not experimentally accessible.  
\begin{figure}
\begin{center}
\includegraphics[angle=-90.,scale=0.5]{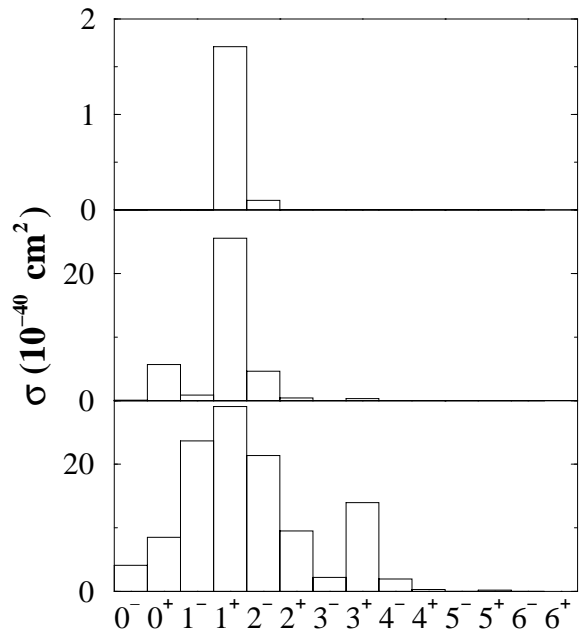}
\end{center}
\protect\caption{{\it Cross section of the $^{208}$Pb$(\nu_{e},e^-)^{208}$B reaction}:
The figure shows how the relative contribution
of states excited in the reaction and having
different multipolarity
increases, for
increasing neutrino energy, namely
for $E_{\nu}=15~$MeV (up), $30~$MeV (middle),
$50~$MeV (bottom) \protect\cite{volpelead}.
Note that little experimental information is available on the
$J^{\pi}$=$0^{-},1^{-},2^{-}$ states and none on those
having higher multipolarity.}
\end{figure}

In order to address these (and other) open questions, one needs a facility
producing intense low-energy neutrinos. 
Note that the Miner$\nu$a proposal
would address interesting issues on neutrino-nucleus
interactions with energetic neutrinos \cite{minerva}.

\section{Strategies}
There are essentially two options for a facility producing
low-energy neutrino beams: either a conventional one, based
on the decay of pions and muons, or beta-beams \cite{lownu}.
The former was the object a few years ago of a proposal, i.e.
ORLAND (Oak Ridge LAboratory for Neutrino Detectors) \cite{orland}
and is now taking a new shape \cite{efremenko}. The latter is a
recent proposition \cite{lownu}, based on a novel method to produce
neutrino beams, which exploits the acceleration of nuclei
that decay through beta-decay \cite{zucchelli}. The neutrino beams
 obtained with these two options present
complementary features both for the flavour content
and for the energy. Conventional sources
provide us with neutrinos of different flavours, while
beta-beams produce pure electron neutrino (or anti-neutrino)
beams. As far as the energy is concerned, muon decay
furnish neutrinos with a Michel spectrum peaked at about
35 MeV and having maximum energy of about 50 MeV.
Beta-beams have the specific feature that the neutrino
mean energy depends on the ion acceleration according
to $E_{\nu}\simeq 2 \gamma Q_{\beta}$,
where $\gamma$ is the Lorentz gamma factor and
$Q_{\beta}$ is the beta-decay Q-value. Therefore the
neutrino energy range can be varied by varying the $\gamma$
of the decaying ions. A detailed comparison for the specific
case of the lead is made in \cite{gailnew}.

\subsection{Low-energy beta-beams}
Two possible scenarios can be envisaged for a low-energy beta-beam
facility \cite{lownu} where it is part either of one of the future nuclear laboratories
(or projects) for the
production very intense exotic ion beams (like e.g. GSI, GANIL, EURISOL or RIA), 
or of the high energy beta-beam
facility at CERN.
For the former case, there are several requirements which
should be met. Two essential aspects are the ion intensity which can be
attained and the availability of a ring to store the ions. 
Let us mention a few cases as typical examples. The future GSI facility 
\cite{gsi} 
includes a storage ring and the ions will be accelerated to GeV energies,
producing neutrinos having a few tens of MeV. However, the fragmentation method
 used to produce the ions will give at maximum $10^{9}~\nu$/s.
At GANIL, the ISOLDE technique employed 
will allow to reach $10^{12}~\nu$/s but no storage ring is planned, 
so that the ions can be eventually used as a neutrino source
at rest.
As far as the high energy beta-beam facility at CERN is concerned,
according to the first baseline scenario,  the beams are
accelerated to several tens of GeV per nucleon
and stored
in a storage ring with long straight sections \cite{zucchelli,mats}. The beams
are fired to a gigantic Cherenkov detector \cite{uno},
located in an upgraded Fr\'ejus Underground
Laboratory, with the aim of studying
very small values of the neutrino mixing angle
$\theta_{13}$ and CP (and T) violation  in the lepton sector
\cite{zucchelli,mauro}. 
Other interesting scenarios are now proposed 
where beta-beams would have even higher energies 
and would be sent to further distances, like e.g. the Gran
Sasso Laboratory \cite{jj,tmms}.
If such a facility is built,
low-energy neutrino beams would be available and could
be fired to a detector located close to the storage ring.

It is important to emphasize that a rich physics
program can be performed, if such beams are available
\cite{jpg,lownu,orland}, neutrino-nucleus interaction studies being one
of the possible axis of research \cite{lownu,julien}. Here we also describe the
potential as far as the neutrino magnetic moment is concerned \cite{munu}.

\section{Neutrino-nucleus interaction rates}

\begin{figure}
\begin{center}
\includegraphics[angle=-90.,scale=0.5]{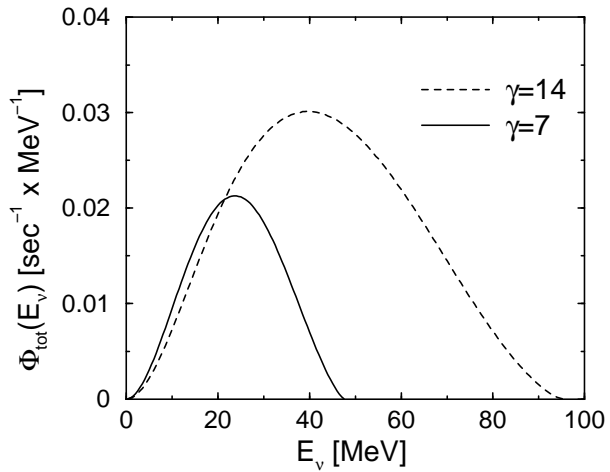}
\end{center}
\caption{{\it Neutrino fluxes at a low-energy
beta-beam facility \cite{julien}:} 
The results shown are obtained using Eq.(\ref{phitot}-\ref{f}) and
correspond to $^{18}$Ne, as a beta-emitter, boosted to
two different Lorentz factors. }
\label{fig:toosmall}
\end{figure}

We present the rates that can be attained at a low-energy
beta-beam facility. 
The total number of events per unit time is
given by \cite{julien}: 
\beq
\label{dNevdt}
 \frac{dN_{ev}}{dt}=g\tau nh\times
 \int_0^\infty dE_\nu\,\Phi_{tot}(E_\nu)\,\sigma(E_\nu)\,,
\eeq
where $n$ is the number of target nuclei per unit volume,
$g$  is the number of
injected ions per unit time,
$\tau$ the half-life of the
parent nucleus,
$\sigma(E_\nu)$
the relevant neutrino-nucleus interaction cross-section as a function
of neutrino energy.  The
neutrino flux $\Phi_{tot}(E_{\nu})$
is obtained by integrating over the useful decay path of
the storage ring and
over the volume of the detector:
\beq
\label{phitot}
 \Phi_{tot}(E_\nu)=\int_0^D \frac{d\ell}{L} \int_0^h \frac{dz}{h}
 \int_0^{\bar{\theta}} f(\theta)\Phi_{lab}(E_\nu,\theta),
\eeq
with
\beq
\label{theta}
\tan\bar{\theta}(\ell,z)=\frac{R}{d+\ell+z}\,,
\eeq
and 
\beq\label{f}
f(\theta)= \frac{\sin\theta d\theta}{2},
\eeq
where $\theta$  is the angle of emission with respect to
the beam axis, $L$ the
total length of the storage
ring with straight sections
$D$, $R$ is the radius of the
cylindrical detector of
depth $h$ placed at distance
$d$ from the storage ring.
The full expression for the
boosted flux $\Phi_{lab}$ is given in \cite{julien}.
Figure 2 shows the neutrino
fluxes used.

\begin{table*}
\caption{}
\label{table:1}
\newcommand{\m}{\hphantom{$-$}}
\newcommand{\cc}[1]{\multicolumn{1}{c}{#1}}
\renewcommand{\tabcolsep}{2pc} 
\renewcommand{\arraystretch}{1.2} 
\begin{tabular}{@{}lllll}
\hline
 Reaction          &  Ref.            & Mass  & Small Ring & Large Ring \\
                      & & (tons) & ($L$=450 m, $D$= 150 m) & ($L$=7 km, $D$=2.5 km) \\   
  \hline
  $\nu +$D          &\cite{revuekubo}  &  \m35   &  \m2363      & \m180
  \\
   $\bar\nu +$D      &\cite{revuekubo}  &  \m35   &  \m25779     &
    \m1956      \\
      $\nu + ^{16}$O    &\cite{nuO}        &  \m952  &  \m6054      &
        \m734       \\
	   $\bar\nu +^{16}$O &\cite{nuO}        &  \m952  &  \m82645     &
	      \m9453      \\
	      $\nu +^{56}$Fe    &\cite{nuFe}       &  \m250  &  \m20768 &
	          \m1611      \\
		       $\nu +^{208}$Pb   &\cite{nuPb}       &  \m360  &  \m103707
		            &  \m7922      \\
			    \hline
			    \end{tabular}\\[2pt]
{\it Neutrino-nucleus interaction rates (events/year) at a low-energy
beta-beam facility \cite{julien}:}
Rates on deuteron, oxygen, iron and lead are shown as examples.
The rates are obtained using Eqs.(\ref{dNevdt}-\ref{f}) with $\gamma=14$
as boost of the parent ion. The neutrino-nucleus cross sections are taken
from referred references.
The detectors are located at 10 meters from the storage ring and have cylindrical
shapes ($R$=1.5 m and $h$=4.5 m for deuteron, iron and lead,
$R$=4.5 m and $h$= 15 m for oxygen, where $R$ is the radius and $h$ is the
depth of the detector).
Their mass is indicated in the second column. Rates obtained for two 
different
storage ring sizes are presented ($L$ is the total
length and $D$ is the length of the straight sections). Here 1 year = $3.2
\times 10^{7}$ s. 
\end{table*}

Table~\ref{table:1} presents the results obtained for four nuclei
as typical examples, i.e. deuteron,
oxygen, iron and lead. Note that
the rates shown are obtained by considering realistic ion intensities, namely
$2 \times 10^{13}~\bar{\nu}$/s (from $^{6}$He decay) and $8 \times 10^{11}~\nu$/s (from
$^{18}$Ne decay), as obtained in the first feasibility study \cite{mats}.
An efficiency of $100 \%$ is considered for all cases.
Final rates will
be given by detailed simulations of
the detector response taking into account possible
backgrounds.
The differences in the $\nu$ and $\bar{\nu}$ rates are due both to the
cross sections and to the different ion intensities at production.

In order to show how the
number of events changes as
a function of the storage
ring length,
two scenarios are envisaged,
where the detector is placed
close either to
a small or to a large
storage ring. We take as
typical sizes those of the ring
planned for the future GSI facility \cite{gsi}, and of
the one considered in the beta-beam baseline
scenario at CERN \cite{mats}.
Note that in \cite{julien} an analytical formula is given which
allows one to scale the present exact
rates for storage rings of different
lengths. In fact, for a close detector as is the case here,
the rates do not simply scale as $L/D$, like for a far detector,
due to the anisotropy of the flux.

From Table 1 one can see that interesting interaction rates can
be achieved by using typical parameters available from
existing feasibility studies.

\subsection{Prospects for the neutrino magnetic moment}
\noindent
The indirect evidence that neutrinos are massive
particles, provided by oscillation experiments,
implies that 
neutrinos have a small magnetic
moment. 
\begin{figure}[h!]
\begin{center}
\includegraphics[scale=0.6, angle=0]{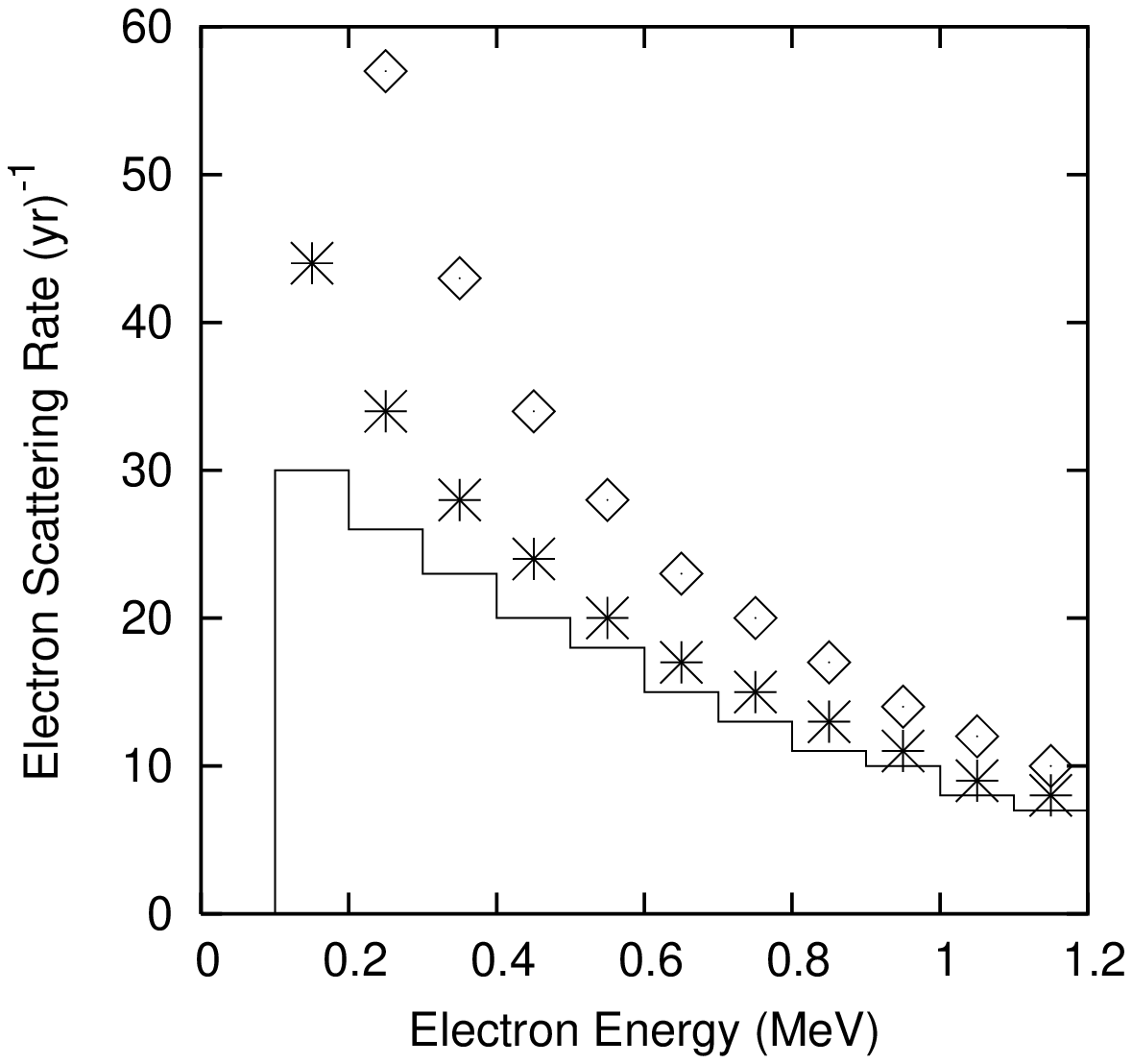}
\end{center}
\caption{\small {\it Prospects on the neutrino magnetic moment \cite{munu}:}
The figure shows the number of neutrino-electron scattering events.
$^{6}$He is the beta-beam emitter
produced at the rate $10^{15}$ per second, and collected inside a
$4 \pi$ detector.
(Similar results are obtained if $^{18}$Ne is used instead.)
The results shown corresponding to electron recoil energies of 
[0.1,1.2] MeV are shown. 
The diamonds show
the number of scatterings if the neutrino has a magnetic moment of
$\mu_\nu = 10^{-10} \mu_B$, the stars present the number of events
if $\mu_\nu = 5 \times 10^{-11} \mu_B$.
The histogram shows the expected number of
events for a vanishing neutrino magnetic moment.
\label{he6-his-kev}}
\end{figure}
\begin{figure}[h!]
\begin{center}
\includegraphics[width=7cm,angle=0]{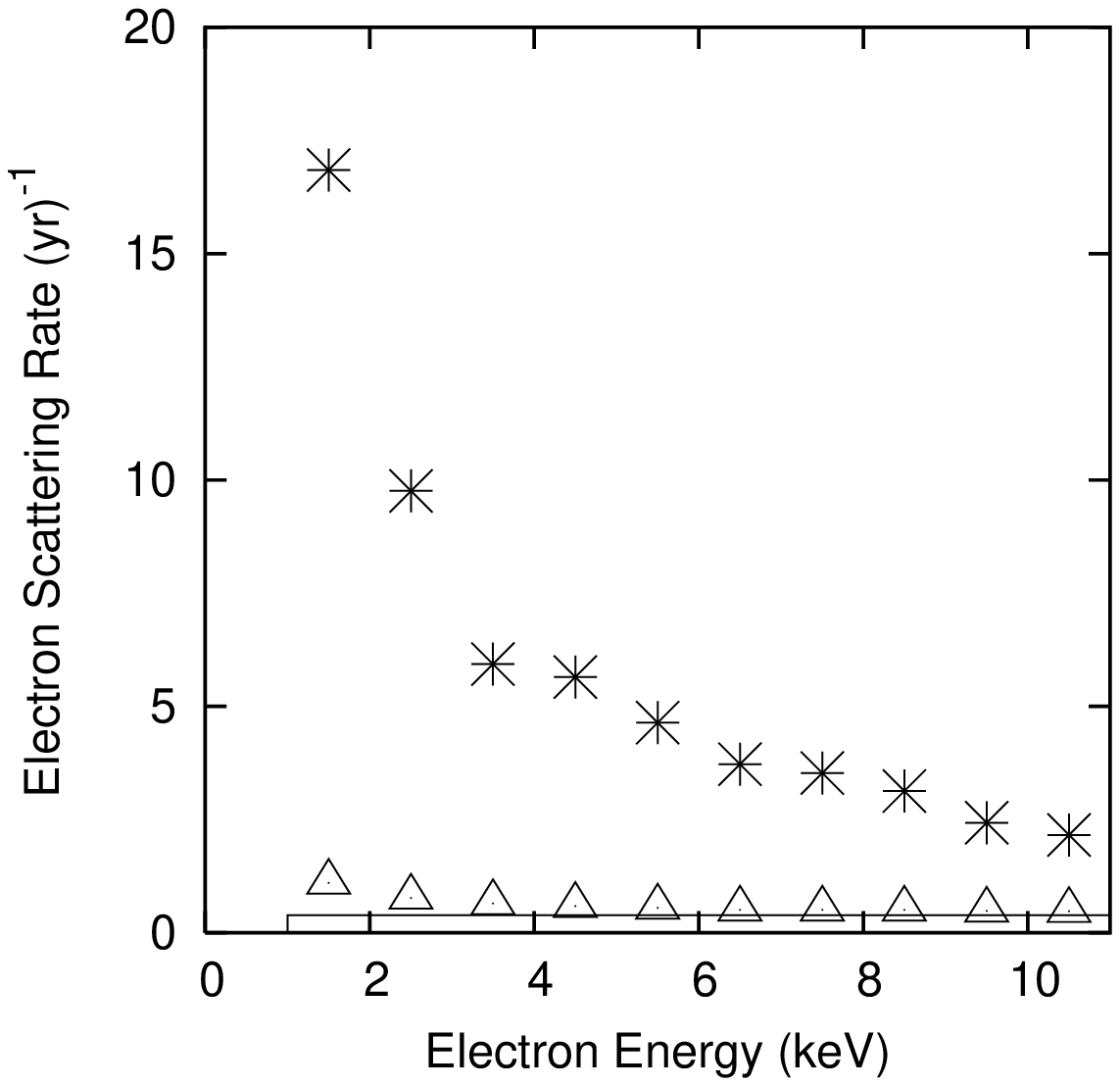}
\end{center}
\caption{\small Same as Figure 1 but for electron energy recoils
within [1,10] keV. The triangles give the number of events
if the neutrino has a magnetic moment of $\mu_\nu = 10^{-11} \mu_B$.
}\label{he6-his-kev}
\end{figure}
In the case of a Dirac mass, standard model interactions
give the neutrino a magnetic moment of
$3 \times 10^{-19} (m_{\nu}/ {\rm eV})$ in units of Bohr
magnetons, $\mu_B$.
The observation of a large magnetic moment would indicate
interactions 
beyond the Standard Model and provide 
valuable information for understanding
the neutrino mass mechanism.
So far, the best limits from direct measurements have been obtained
with reactor experiments and are in the range
$\mu_{\nu} < 1.0-4 \times 10^{-10} \mu_B$ at $90$~\% C.L.
\cite{Daraktchieva:2003dr}. 
Similar upper bounds have recently been deduced from
solar events \cite{skmunu}.
Indirect limits
in the range $10^{-11}- 10^{-12} \mu_B$
have been obtained
by using astrophysical considerations \cite{Raffelt:wa},
although the exact values for these
limits are model-dependent (for a review see \cite{wongcont}).

The direct measurements 
exploit neutrino-electron scattering where the neutrinos
are detected by measuring the recoil of the electrons.
In fact, if the magnetic moment is non-zero
an extra electromagnetic term adds to the
cross section \cite{Daraktchieva:2003dr}:
\begin{equation}
\left({d\sigma \over dT}\right)_{\sc M} =
{\pi \alpha^2 \mu_{\nu}^2 \over{m_e^2}} {1-T/{E_{\nu}} \over T} ,
\end{equation}
where 
$T$ is the electron recoil energy, 
 $m_e$ is the electron mass.
One can see that
a non-zero neutrino magnetic moment dominates the
neutrino-electron cross section
particularly for very low electron recoils ($T \rightarrow 0$).
This fact is exploited in direct measurements
to set a limit on $\mu_{\nu}$.

Here we present the potential of a low-energy beta-beams facility \cite{munu}.
The ions are used as an intense neutrino source at rest.
To improve present direct limits on the neutrino magnetic moment
 one needs :
{\it i)} very intense neutrino sources of well-known fluxes;
{\it ii)} very low threshold detectors.
Such detectors are currently investigated
\cite{mamont}.
Clearly in this case - as for a static source 
\cite{mamont,wong,ioannis} -
the neutrino fluxes can be very accurately calculated.
Figures 3 and 4 show the number of neutrino-electron scattering events
as a function of electron energy recoil.
The results are obtained by averaging
the total (weak and electromagnetic) cross section
with the neutrino fluxes produced by collecting
$10^{15}~$ $^{6}$He/s inside a 4$\pi$ detector
(such intensities might be attained with further feasibility studies
\cite{matscom}).
A $100 \%$ efficiency is assumed.
If there is no magnetic moment, this intensity
will produce about 170
events in the 0.1 MeV to 1 MeV range
per year and  3 events in the 1 keV to 10 keV range per year.
These numbers increase to 210 and 55 respectively in the case of a
magnetic moment of  $5 \times 10^{-11} \mu_B$.
This indicates that the present direct limits might be improved by
 almost an order of magnitude, the precise value requiring
 a detailed simulation of the detector response.

\section{Conclusions and Perspectives}

The availability of low-energy neutrino beams
would offer the opportunity to tackle interesting open
issues on neutrino-nucleus interactions of interest
for various domain of physics. The option of a low-energy beta-beam
facility seems a very promising one. 
For these low energy applications it would be of great 
interest to investigate if, at least for one beta-beam emitter,
higher ion production rates, than the ones obtained in the first feasibility
study,
can be achieved. In the coming years
a detailed feasibility study of the low-energy as well as 
the high-energy beta-beam facility will be performed
within the context of the 
European Isotope Separation On-Line 
Radioactive Ion Beam Facility (EURISOL) project.


\begin{thebibliography}{9}
\bibitem{gail}   G.McLaughlin and G.M. Fuller, Astrophys. J. 455 (1995)
202.
\bibitem{qian} Y.Z. Qian et al, Phys. Rev. C 55 
  (1997) 1532.
\bibitem{goriely} I.N. Borzov and S. Goriely, Phys. Rev. C 62 (2000) 035501-1.
\bibitem{bahanunucl} A.B. Balantekin, Prog. Theor. Phys.
Suppl. 146 (2003) 227 [nucl-th/0201037].
\bibitem{bahaastro}  A.B. Balantekin and G.M. Fuller, J. Phys. G 29
(2003) 2513
[astro-ph/0309519].
\bibitem{terasawa} M.Terasawa et al, Astrophys. J. 608 (2004) 470. 
\bibitem{haxton} S.E. Woosley et al, Astrophys. J. 356 (1990) 272.
\bibitem{woosley} A. Heger et al, astro-ph/0307546.
\bibitem{haxtonrec} W.C. Haxton, nucl-th/0406012.
\bibitem{pinedo} K. Langanke and G. Martinez-Pinedo, Nucl. Phys. A 731 (2004) 365. 
\bibitem{deut} S.E. Willis et al, Phys. Rev. Lett. D 4
(1980) 522.
\bibitem{iron} E. Kolbe, K. Langanke and G. Martinez-Pinedo, Phys. Rev. C
60 (1999) 052801.
\bibitem{expc12} C. Athanassopoulos and the LSND collaboration,
Phys. Rev. C  56 2806 (1997) 2806; M. Albert et al,
Phys. Rev. C  51 (1995) R1065; C. Athanassopoulos and the LSND
collaboration,
Phys. Rev. C 55 (1997) 2078; D.A. Krauker et al, Phys. Rev. C 
45 (1992) 2450;
  R.C. Allen  et al,
      Phys. Rev. Lett. 64 (1990) 1871; B.E. Bodmann and the KARMEN
          collaboration,
	        Phys. Lett. B332 (1994) 251; J. Kleinfeller 
		 in Neutrino 96, eds. K.Enquist, H.Huitu and J.Maalampi
		            (World Scientific Singapore, 1997).
\bibitem{revuekubo} K. Kubodera and S. Nozawa, Int. J. Mod. Phys. E
3 (1994) 101 and references therein.
\bibitem{jpg} See also J. of Phys. G 29  (2003) 2497.

\bibitem{sno} The SNO Collaboration, Phys. Rev. Lett.  87 
(2001) 071301  and
Phys. Rev. Lett. 89 (2002) 011301.

\bibitem{kubodera} S. Ying, W. C. Haxton, and E. M. Henley,
Phys. Rev. C 45 (1992) 1982;
D.B. Kaplan, M.J. Savage, M.B. Wise, Phys. Lett. B 424 (1998) 390;
M. Butler, J.-W.Chen, X. Kong,  Phys. Rev. C 63
 (2001) 035501 [nucl-th/0008032];
K. Kubodera, Nucl. Phys. Proc. Suppl. 100  (2001) 30;
M. Butler, J.-W. Chen, P. Vogel, Phys. Lett. B 549  (2002) 26
[nucl-th/0206026];
A.B. Balantekin and H. Y\"uksel, hep-ph/0307227.
\bibitem{c12} E. Kolbe et al,
  Phys. Rev. C 52 (1995) 3437; N. Auerbach, N. Van Giai and O.K. Vorov,
    Phys. Rev. C 56 (1997) R2368; S.K. Singh, N.C. Mukhopadyhay and E.
      Oset, Phys. Rev. C 57  (1998) 2687; S.L. Mintz and M. Pourkaviani,
        Nucl. Phys. A 594  (1995) 346; E. Kolbe, K. Langanke and
	  P. Vogel, Nucl. Phys. A 613 (1997) 382;
	    A.C. Hayes and I.S. Towner, Phys. Rev. C 61 (2000) 044603;
	    C. Volpe et al, Phys. Rev. C 62  (2000) 015501;
	    N. Auerbach and B.A. Brown, Phys. Rev. C 65 (2002) 024322;
	      N. Jachowicz et al, Phys. Rev. C 65 (2002) 025501.
\bibitem{uno} C.K. Jung, Proceedings of the Next generation Nucleon Decay
and Neutrino Detector (NNN99) Workshop, September 23-25, 1999, Stony
Brook, New York [hep-ex/0005046].
\bibitem{fuller} G.M. Fuller, W.C. Haxton and G.C. McLaughlin,
Phys. Rev. D 59 (1999) 085005.
\bibitem{elliott}
N. Jachowicz, K. Heyde and J. Ryckebush, Phys. Rev. C 66 (2002) 055501;
S.R. Elliott, Phys. Rev. C 62 (2000) 065802;
P.F. Smith, Astropart. Phys. 8 (1997) 27;
C.K. Hargrove et al, Astropart. Phys. 5 (1996) 183;
D.B. Cline et al, Phys. Rev. D. 50 (1994) 720.
\bibitem{volpelead} C. Volpe et al, Phys. Rev. C 65
(2002) 044603.
\bibitem{nuPb} J. Engel, G.C. McLaughlin,
C. Volpe, Phys. Rev. D 67 (2003) 013005.
\bibitem{nuFe} E. Kolbe and K. Langanke, Phys. Rev. C 63 (2001) 025802.
\bibitem{kolbe} E. Kolbe et al, Nucl. Phys. A 540  (1992) 599.
\bibitem{jon} R. Surman and J. Engel, Phys. Rev. C 58 (1998) 2526.
\bibitem{minerva} See contribution of J. Morfin to this volume and 
http://www.pas.rochester.edu/~ksmcf/minerva/.
\bibitem{lownu} C. Volpe, Jour. Phys. G 30 (2004) L1
 [hep-ph/0303222].
 \bibitem{orland} F.T. Avignone et al,
 Phys. Atom. Nucl. 63  (2000) 1007; see http://www.phy.ornl.gov/orland/.
\bibitem{efremenko} Y. Efremenko, private communication.
\bibitem{zucchelli} P. Zucchelli, Phys. Lett. B  532 (2002) 166.
\bibitem{gailnew} G.C. McLaughlin, nucl-th/0404002.
 \bibitem{gsi} See http://www.gsi.de/.
\bibitem{mats} B. Autin et al, J. Phys. G 29  (2003) 1785
 [physics/0306106]; M. Lindroos, physics/0312042.
 See also http://beta-beam.web.cern.ch/beta-beam/.
 \bibitem{mauro} M. Mezzetto, J. Phys. G 29  (2003) 1771
 [hep-ex/0302007]; J. Bouchez, M. Lindroos, M. Mezzetto,
 Proceedings to Nufact03 [hep-ex/0310059];
 see contribution of M. Mezzetto to this volume.
\bibitem{jj} J. Burguet-Castell et al IFIC/03-55 [hep-ph/0312068].
 \bibitem{tmms} F. Terranova, A. Marotta, P. Migliozzi, M. Spinetti,
 hep-ph/0405081.
 \bibitem{julien} J. Serreau and C. Volpe, submitted for publication
    [hep-ph/0403293].   
\bibitem{munu} G.C. McLaughlin and C. Volpe,
 Phys. Lett. B  591 (2004) 229 [hep-ph/0312156].
\bibitem{nuO} E. Kolbe, K. Langanke, P. Vogel, Phys. Rev. D 66 
(2002) 013007; W.C. Haxton, Phys. Rev. D 36 (1987) 2283.
\bibitem{Daraktchieva:2003dr}
Z.~Daraktchieva et al  [MUNU Collaboration],
Phys.\ Lett.\ B 564  (2003) 190
[hep-ex/0304011];  H.B. Li, et al, TEXONO Collaboration,
Phys. Rev. Lett. 90  (2003) 131802;
F.~Reines, H.S.~Gurr, and H.W. Sobel,
Phys. Rev. Lett. 37  (1976) 315;
P.~Vogel and J.~Engel, Phys.\ Rev.\ D 39 (1989) 3378;
G.S.~Vidyakin {\it et al.}, JETP Lett.\  55 (1992) 206
[Pisma Zh.\ Eksp.\ Teor.\ Fiz.\  55 (1992) 212];
A.~I.~Derbin, A.~V.~Chernyi, L.~A.
~Popeko, V.~N.~Muratova, G.~A.~Shishkina and S.~I.~Bakhlanov,
JETP Lett.\  57 (1993) 768
[Pisma Zh.\ Eksp.\ Teor.\ Fiz.\  57  (1993) 755].
\bibitem{skmunu} The Super-Kamiokande Collaboration, Phys. Rev. Lett. 
93  (2004) 021802; M.A. Tortola, Proceeding to International Workshop on
Astroparticle and High-Energy Physics, Valencia, 2003 [hep-ph/0401135];
O.G. Miranda et al, Phys. Rev. Lett. 93 (2004) 051304;
J.F. Beacom and P. Vogel, Phys. Rev. Lett. 83 (1999) 5222.
\bibitem{Raffelt:wa}
G.~G.~Raffelt,
Stars As Laboratories For Fundamental Physics: The Astrophysics Of
Neutrinos, Axions, And Other Weakly Interacting Particles,
Chicago, USA: Univ. Pr. (1996); and references therein.
\bibitem{wongcont} See contribution of H. Wong to this volume
 [hep-ex/0409003].
\bibitem{mamont} The Mamont Collaboration, Nucl. Phys. A
721 (2003) 499.
\bibitem{wong}
H.-B. Li and H.T. Wong, J. Phys. G. 28  (2002) 1453;
V.N. Trofimov et al, Phys. At. Nuclei 61,
1271 (1998) 1271.
\bibitem{ioannis} Y. Giomataris and J. Vergados, hep-ex/0303045.
\bibitem{matscom} M. Lindroos, private communication.

\end{thebibliography}
\end{document}